\documentclass[draft,preprint,aps,
showpacs
]{revtex4}

\usepackage{amsmath}
\usepackage{amssymb}
\usepackage{graphicx}
\usepackage{bm}
\usepackage{epsfig}
\input{epsf}
\begin{document}

\preprint{Guchi-TP-022}

\title{
Finite density effects in Hosotani mechanism and \\
a vacuum gauge ball
}

\author{Yoshinori Cho}
\email{b2669@sty.cc.yamaguchi-u.ac.jp}
\affiliation{
Graduate School of Science and Engineering, Yamaguchi University
\\
Yoshida, Yamaguchi-shi, Yamaguchi 753-8512, Japan
}
\author{Kiyoshi Shiraishi}
\email{shiraish@sci.yamaguchi-u.ac.jp}
\affiliation{
Faculty of Science, Yamaguchi University
\\ Yoshida, Yamaguchi-shi, Yamaguchi 753-8512, Japan
}

\date{\today}
\begin{abstract}
We consider the finite density effects of the fermion
with $U(1)$ gauge symmetry in Hosotani mechanism.
We construct a vacuum gauge ball,
a new kind of non-topological soliton, and
investigate their properties numerically.
We find the relations between the physical quantities.
\end{abstract}
\maketitle
\section{Introduction}
Recently, there has been much interest in symmetry breaking
mechanisms in theories with extra dimensions.
(For a recent review, see Ref. \cite{ref1}.)
In particular, Hosotani mechanism \cite{ref2,ref}
attracts much attention and has been studied by many authors.
In this mechanism, the vacuum expectation value (VEV) of the
gauge field configuration on the multiply-connected space,
which is related with the Wilson line element
on the space, plays an important role.
The Wilson line elements on the space become dynamical
degrees of freedom and parametrize degenerate classical vacua.
The quantum effects (in terms of the one-loop effective potential)
lift the degeneracy and determine the physical vacua.
If the effective potential has a minimum
corresponding to a non-trivial configuration,
the symmetry is spontaneously broken or enhanced.

On the other hands, in cosmological context,
the analysis of high temperature or high density effects
is essential to understand the phase transition during
the cosmological evolution.
In ordinary field theories including spontaneous symmetry
breaking, the symmetry is expected to be restored
at high temperature \cite{linde} or high density \cite{bl}.
In Hosotani mechanism, the finite density and
high temperature effects have been investigated
in \cite{ref3,ref4,ref5,ref6,ref7}.
One of the authors found that the symmetry
can be broken or restored by high density effects
of the degenerate fermion \cite{ref3}.
The various ``phases" may exist inside an object of high density.

In this paper, we consider the finite density effects
for the degenerate fermion in Hosotani mechanism.
Instead of studying breaking and restoration of symmetry
by the effects, we construct an object
by similar methods to Fermi ball (F-ball) \cite{Fball1,Fball2},
and Q-ball \cite{Qball} such as a kind of
non-topological soliton (NTS) \cite{NTS},
which may be related to the important clue to solve problems
in cosmology, for example, dark matter problem.
We call the new object ``vacuum gauge ball".

NTS is formed as a stable solution whose boundary
condition at infinity is expressed as the vacuum state.
Q-ball, which is a kind of NTS,
is stabilized by conservation of global $U(1)$ charge.
One of the characters of the scalar potential associated with them
is that there is a barrier between the symmetric and broken phases.
The conceptual construction of F-ball,
on the other hands, consists of three steps.
The first step is that there are two almost degenerate vacua
(one is true vacuum and another is false one),
which appear in the spontaneous breaking
of an approximate $Z_2$ symmetry, which is called
``biased $Z_2$ symmetry" in \cite{Fball1}.
At the phase transition,
a domain wall is produced between two vacua
and zero-mode fermions are captured in the domain wall.
The next is that if the true vacuum is energetically
favored, then a region of false vacuum gets diminished
and continues to shrink due to the surface energy 
as well as the volume energy inside the domain wall.
Finally, the Fermi energy stops the shrinkage
and keeps the dynamical balance.
Such a bag of the false vacuum with zero-mode
fermions caught in the surrounding wall
is called F-ball.

If Hosotani mechanism is applied to the case of finite density,
it is also possible to realize the similar situations to
the case of F-ball and Q-ball:
the thermodynamic potential (or the effective potential)
in Hosotani mechanism
has a barrier between their vacua, appeared in case of Q-ball.
Further, the two vacua are almost degenerate with the
energy density difference, parametrized by
(chemical potential) $\times$
(circumference of the extra dimension).
When the vacuum gauge field forms a domain wall
configuration, the vacuum gauge ball is constructed,
owing to the dynamical balance between the shrinking force
due to the energy difference of the two vacua
and the expanding force due to the Fermi energy,
appeared in such case as F-ball.

In this setting, we investigate the relationship
between the mass, the radius and the particle number
of the vacuum gauge ball
in $M_3\times S^1$ and $M_4\times S^1$ space-times.

The organization of this paper is as follows.
In the next section, in order to consider the finite density effects
in Hosotani mechanism, we evaluate the thermodynamic potential
for the vacuum gauge field, using the zeta function regularization
technique.
We construct a vacuum gauge ball and investigate their property
numerically in $M_3\times S^1$ and $M_4\times S^1$ space-times
in section III.
Finally, we devote section IV to discussion.
\section{Finite density effects in Hosotani Mechanism}
In this section, we begin by considering a gauge theory
with a Dirac fermion in $M_{(d+1)}\times S^1$ space-time.
Let $x^{i}$ $(i=0,1,\dots,d)$ and $y$
label the $(d+1)$-dimensional Minkowski space-time
and $S^1$ space whose circumference is $L$ $(0\leq y<L)$,
respectively, ($\mu$, $\nu$ run over both $i$ and $y$).
In our model, we treat $U(1)$ gauge symmetry, for simplicity.
Here we still call the mechanism (of determining the non-trivial
Wilson element by quantum effects) as Hosotani mechanism,
even if the symmetry is abelian (and is not broken).
Generalizing to the cases of various gauge symmetries
and topology of the extra dimensions are straightforward.

The Lagrangian is denoted by
\begin{equation}
 {\cal L}=-\frac{1}{4}F_{\mu\nu}F^{\mu\nu}
 +i\bar{\psi}\gamma^{\mu}D_{\mu}\psi,
 \label{lag}
\end{equation}
where the Maxwell field strength and the covariant derivative
with the gauge coupling constant $g$ are
given by
\begin{eqnarray}
 F_{\mu\nu}&=&\partial_{\mu}A_{\nu}-\partial_{\nu}A_{\mu}, \\
 D_{\mu}&=&\partial_{\mu}-igA_{\mu},
\end{eqnarray}
respectively.
Since $S^1$ is a non-simply connected space,
one must specify the boundary conditions of the fields.
The boundary conditions on the gauge field
and the fermion field are chosen as follows
\begin{eqnarray}
 A_{\mu}(x,y+L)&=&A_{\mu}(x,y), \\
 \psi(x,y+L)&=&e^{i\delta}\psi(x,y).
 \label{fermionbc}
\end{eqnarray}
The phase $e^{i\delta}$ cancels from physical operators
of a bilinear form $\bar{\psi}\psi$ but contributes to
the boundary conditions.

In $S^1$ space,
non-zero vacuum gauge field configuration is admitted.
Then, we set the VEV of the gauge field configuration
\begin{equation}
 \phi=\langle g L A_y \rangle,
\end{equation}
where $A_y$ denotes a component of the gauge field on $S^1$.
If we just consider non-singular gauge transformation,
non-zero $\langle g L A_y \rangle$
is inequivarent to the trivial vacuum gauge field
$\langle g L A_y \rangle=0$ unless
$\langle g L A_y \rangle=2\pi n$ ($n$: integer).
Then one can only restrict $0\leq \phi <2\pi$
by gauge transformations.
Note that the VEV of the gauge field configuration
$\phi$ plays the role of an ``order parameter"
and the quantum effects in terms of the effective potentials
determine the order parameter.
For instance, one finds the expression of
the one-loop effective potential for the vacuum gauge field
from the massless Dirac fermions
\begin{equation}
 V_{\mathrm{eff}}=
 -({\rm tr}{\bf 1})\frac{1}{2}\frac{N_f}{(LV_d)}
 \ln\det(D^2),
 \label{eff}
\end{equation}
where $V_d$ is the volume of $d$-dimensional space.
$N_f$ and ${\rm tr}{\bf 1}(=2^{[(d+2)/2]}$) are
the number of a Dirac fermion and the number of
components of the fermion, respectively.
$[x]$ is the integral part of $x$.
If the effective potential for $\phi$ is  minimized at
a non-trivial configuration, then the fermions acquire
dynamical masses from the quantum effects.
The effective potential (\ref{eff}) can be evaluated
by the zeta function regularization method \cite{ref12,ref14,ref15},
\begin{equation}
 \ln\det(D^2)= -\zeta'(0),
\end{equation}
where $\zeta(s)$ is the zeta function, which is defined
according to the field contents.

Now, we shall consider the finite density effects
in Hosotani mechanism.
For this purpose, we adopt the usual imaginary time formalism,
according to the standard techniques in finite-temperature
field theory \cite{ref13}:
the time coordinate is Wick rotated to the
Euclidean time $\tau=it$ $(0\leq \tau <\beta)$
with boundary conditions on fields
(periodic for bosons and anti-periodic for fermions).
The chemical potential $\mu$ of the fermion field
is introduced as the zeroth component of the covariant derivative
to be modified by
\begin{equation}
 D_0=\partial_0 \to \partial_0-iA_0,
 \qquad
 \mbox{where} \quad A_0=-i\mu.
\end{equation}

Employing the zeta function regularization method,
the effective potential, or
the thermodynamic potential $\Omega$ in statistical mechanics
is expressed as the following form
at temperature $\beta^{-1}$ and chemical potential $\mu$:
\begin{equation}
 \beta\Omega=({\rm tr}{\bf 1})\frac{1}{2}\frac{N_f}{LV_d}
 \zeta'(0),
\end{equation}
with
\begin{eqnarray}
 \zeta(s)
 &=&
 \frac{2V_d}{\Gamma(s)}
 \int^{\infty}_0dtt^{s-1}
 \int\frac{d^dk}{(2\pi)^d}
 \nonumber \\
 & & \times
 \sum^{\infty}_{n=-\infty}\sum^{\infty}_{l=-\infty}
 \exp\left[
 -t\left\{
 k^2+\left(\frac{(2n+1)\pi}{\beta}+i\mu\right)^2
 +\left(\frac{2\pi l+\phi}{L}\right)^2
 \right\}
 \right],
 \label{eq3}
\end{eqnarray}
where, for simplicity, we have restricted the Dirac fermions
to the case of periodic boundary condition on $S^1$, {\it i.e.},
$\delta=0$ in (\ref{fermionbc}).
The factor ``two" of the right hand side in (\ref{eq3}) is
the contributions from the degeneracy of the fermion
in the case that the compact manifold is $S^1$.
In general, the degeneracy of the fermion is given by
\begin{equation}
 d_l(N)=\frac{2\Gamma(N+l)}{l!\Gamma(N)},
\end{equation}
for a sphere $S^N$ case \cite{cw}.

In order to carry out the calculation,
it is convenient to use the following identity
in terms of theta function
\footnote{For the explanation of the identity,
see the Appendix in Ref. \cite{ref14}.}
\begin{equation}
 \sum^{\infty}_{n=-\infty}
 \exp
 \left[
 -t
 \left(
 \frac{(2n+1)\pi}{\beta}+i\mu
 \right)^2 
 \right]
 =
 \frac{\beta}{(4\pi t)^{1/2}}
 \left[
 1+2\sum^{\infty}_{n=1}(-1)^n
 \cosh(n\beta\mu)
 \exp\left(
 -\frac{\beta^2n^2}{4t}
 \right)
 \right].
\end{equation}
Then one can divide $\zeta(s)$ into two parts
\begin{eqnarray}
 \zeta(s)&=&
 \zeta_0(s)+\zeta_{\beta}(s),
 \\
 \zeta_0(s)&=&
 \frac{2\beta}{(4\pi)^{1/2}}
 \frac{V_d}{\Gamma(s)}
 \int^{\infty}_0dtt^{s-\frac{3}{2}}\int\frac{d^dk}{(2\pi)^d}
 \sum^{\infty}_{l=-\infty}\exp
 \left[
 -t
 \left\{
 k^2+\left(\frac{2\pi l+\phi}{L}\right)^2
 \right\}\right],
 \label{zeta0}
 \\
 \zeta_{\beta}(s)&=&
 \frac{4\beta}{(4\pi)^{1/2}}
 \frac{V_d}{\Gamma(s)}
 \int^{\infty}_0dtt^{s-\frac{3}{2}}\int\frac{d^dk}{(2\pi)^d}
 \sum^{\infty}_{l=-\infty}
 \sum^{\infty}_{n=1}(-1)^n\cosh(n\beta\mu)
 \nonumber \\
 & &
 \times\exp
 \left[-t\left\{k^2+\left(\frac{2\pi l+\phi}{L}\right)^2
 \right\}-\frac{\beta^2n^2}{4t}
 \right],
\end{eqnarray}
where $\zeta_0(s)$ part contributes to the thermodynamic potential
as the temperature-independent vacuum energy
and $\zeta_{\beta}(s)$ part is the temperature-dependent
contribution to the vacuum energy.
In the following subsections, we calculate each part
of the zeta function $\zeta(s)$ in turn.
\subsection{calculation of $\zeta_0(s)$ part}
In calculation of $\zeta_0(s)$ part,
one can also use the following identity in terms of
theta function in (\ref{zeta0})
\begin{equation}
 \sum^{\infty}_{l=-\infty}
 \exp\left[-t
 \left(\frac{2\pi l+\phi}{L}\right)^2
 \right]=
 \frac{L}{(4\pi t)^{1/2}}
 \sum^{\infty}_{l=-\infty}
 \exp\left[-\frac{L^2}{4t}l^2\right]
 \cos(l\phi),
\end{equation}
where we have only adopted the real part of the right hand side
in the identity because the imaginary part does not contribute
to the thermodynamic potential.

Then, the thermodynamic potential contributed from
the temperature-independent part,
which is denoted by $\Omega_0$, is reduced to
\begin{eqnarray}
 \Omega_0
 &\equiv&
 \frac{({\rm tr}{\bf 1})}{2\beta}\frac{N_f}{V_dL}\zeta'_0(0),
 \\
 &=&
 ({\rm tr}{\bf 1})
 \frac{N_f}{(4\pi)^{(d+2)/2}}
 \int^{\infty}_0dtt^{-\frac{d+2}{2}-1}
 \sum^{\infty}_{l=-\infty}
 \exp\left[-\frac{L^2}{4t}l^2\right]
 \cos(l\phi),
 \label{17}
 \\
 &=&
 ({\rm tr}{\bf 1})
 \frac{2N_f}{\pi^{\frac{d+2}{2}}}
 \Gamma\left(\frac{d+2}{2}\right)
 \sum^{\infty}_{l=1}\frac{1}{(Ll)^{d+2}}
 \cos(l\phi),
 \label{omegazero}
\end{eqnarray}
where in the second line (\ref{17}), we have performed
the integration of the momentum $k$
and, in the third line (\ref{omegazero}), we have replaced
the summation on $l$ in the Kaluza-Klein modes as follows
\begin{equation}
 \frac{L^2l^2}{4t}\to t,
\end{equation}
which causes the divergence at $l=0$
and have omitted the divergence,
which is the contributions to the cosmological constant.
We assume that the cosmological constant of the true vacuum
one obtains is set to zero by some mechanisms.

In $M_3\times S^1$ space-time, if the $\phi$ is
restricted to the case of $0\leq\phi\leq\pi$,
then the thermodynamic potential
can be obtained as an analytical form
\begin{eqnarray}
 \Omega_0
 &=&
 \frac{8N_f}{\pi^2L^4}
 \sum^{\infty}_{l=1}
 \frac{\cos(l\phi)-(-1)^l}{l^4},
 \\
 &=&
 \frac{N_f}{6}
 \frac{1}{\pi^2L^4}
 \left\{
 2\pi^2
 (\phi-\pi)^2
 -(\phi-\pi)^4
 \right\},
 \label{omegazero2}
\end{eqnarray}
here we have used the useful identities
\begin{equation}
 \sum^{\infty}_{n=1}\frac{\cos nx}{n^4}
 =\frac{1}{48}\left[
 2\pi^2(x-\pi)^2-(x-\pi)^4-\frac{7}{15}\pi^4
 \right],
\end{equation}
and $\sum^{\infty}_{l=1}\frac{(-1)^l}{l^4}=-\frac{7}{8}\zeta(4)$.
\subsection{calculation of $\zeta_{\beta}(s)$ part}
Next, we calculate the thermodynamic potential of
$\zeta_{\beta}(s)$ part as the temperature-dependent
contribution to the vacuum energy,
which is denoted by $\Omega_{\beta}$,
\begin{eqnarray}
 \Omega_{\beta}
 &\equiv&
 ({\rm tr}{\bf 1})
 \frac{1}{2\beta}\frac{N_f}{V_dL}\zeta'_{\beta}(0),
 \\
 &=&
 ({\rm tr}{\bf 1})
 \frac{2}{(4\pi)^{\frac{d+1}{2}}}
 \frac{N_f}{L}
 \int^{\infty}_0dtt^{-\frac{d+1}{2}-1}
 \sum^{\infty}_{l=-\infty}
 \sum^{\infty}_{n=1}(-1)^n\cosh(n\beta\mu)
 \nonumber \\
 & &
 \times
 \exp\left[-t\left(\frac{2\pi l+\phi}{L}\right)^2
 -\frac{\beta^2n^2}{4t}
 \right],
\end{eqnarray}
where we have performed the integration of the momentum $k$.
Moreover we divide three parts for the summation on $l$
($l=0$, plus and minus parts of $l$) and use the following integral
representation of the modified Bessel function \cite{ref16}
\begin{equation}
 K_{\nu}(z)=
 \frac{1}{2}
 \left(
 \frac{z}{2}
 \right)^{\nu}
 \int^{\infty}_0
 \exp\left(
 -t-\frac{z^2}{4t}
 \right)
 t^{-\nu-1}dt.
\end{equation}
Then the thermodynamic potential contributed from
the temperature-dependent part is reduced to
\begin{eqnarray}
 \Omega_{\beta}
 &=&
 ({\rm tr}{\bf 1})
 \frac{4}{(4\pi)^{\frac{d+1}{2}}}
 \frac{N_f}{L}
 \sum^{\infty}_{n=1}
 (-1)^n\cosh(\mu\beta n)
 \left[
 \left(
 \frac{2\phi}{L\beta n}
 \right)^{\frac{d+1}{2}}
 K_{\frac{d+1}{2}}\left(\frac{\beta n}{L}\phi \right)
 \right. 
 \nonumber \\
 & &
 \left.
 +
 \sum^{\infty}_{l=1}
 \left\{
 \left(\frac{2(2\pi l+\phi)}{L\beta n}\right)^{\frac{d+1}{2}}
 K_{\frac{d+1}{2}}
 \left(
 \beta n\frac{(2\pi l+\phi)}{L}
 \right)
 +[\phi\to -\phi]
 \right\}
 \right].
\end{eqnarray}
We use an integral representation of
the modified Bessel function \cite{ref16}
\begin{equation}
 K_{\nu}(z)=
 \frac{\sqrt{\pi}(z/2)^{\nu}}{\Gamma(\nu+1/2)}
 \int^{\infty}_1e^{-zx}(x^2-1)^{\nu-\frac{1}{2}}dx.
\end{equation}
Then one can perform the summation over $n$ to obtain
\begin{eqnarray}
 \Omega_{\beta}
 &=&
 -({\rm tr}{\bf 1})
 \frac{N_f}{L(4\pi)^{\frac{d}{2}}
 \Gamma(\frac{d+2}{2})}
 \int^{\infty}_1dx(x^2-1)^{\frac{d}{2}}
 \left[
 \left(
 \frac{\phi}{L}
 \right)^{d+1}
 \left(
 \frac{1}{\exp[\beta(\frac{\phi x}{L}-\mu)]+1}
 +(\mu\to -\mu)
 \right)
 \right. \nonumber \\ 
 &+&
 \left.
 \sum^{\infty}_{l=1}
 \left\{
 \left(
 \frac{2\pi l+\phi}{L}
 \right)^{d+1}
 \left(
 \frac{1}{\exp\left
 [\beta\left(\frac{(2\pi l+\phi)x}{L}-\mu
 \right)
 \right]
 +1}
 +(\mu\to -\mu)
 \right)
 +[\phi\to -\phi]
 \right\}
 \right].
 \label{eq19}
\end{eqnarray}
Further, in order to describe a system, which consists of
the degenerate fermion gas, {\it i.e.}, in the situation of
$\mu\neq0$ and $T\to 0$,
one can use the fact that,
in the zero-temperature limit $T\to 0$ ($\beta\to\infty$),
\begin{equation}
 \frac{1}{e^{\beta x}+1}
 \stackrel{\beta\to\infty}{\longrightarrow}\theta(-x),
 \label{zero}
\end{equation}
where $\theta(x)$ is the step function.
The expression (\ref{eq19}) reduces to the following form
by means of (\ref{zero}) in the zero-temperature limit
\begin{eqnarray}
 \Omega_{\beta}
 &=&
 -({\rm tr}{\bf 1})
 \frac{N_f}{L(4\pi)^{\frac{d}{2}}
 \Gamma(\frac{d+2}{2})}
 \left[
 \left(
 \frac{\phi}{L}
 \right)^{d+1}
 \int^{\mu/\omega_1}_1
 (x^2-1)^{\frac{d}{2}}dx
 \right. \nonumber \\ 
 &+&
 \left.
 \sum^{l_m}_{l=1}
 \left(
 \frac{2\pi l+\phi}{L}
 \right)^{d+1}
 \int^{\mu/\omega_2}_1
 (x^2-1)^{\frac{d}{2}}dx
 +
 \sum^{l_n}_{l=1}
 \left(
 \frac{2\pi l-\phi}{L}
 \right)^{d+1}
 \int^{\mu/\omega_3}_1
 (x^2-1)^{\frac{d}{2}}dx
 \right],
\end{eqnarray}
where $\omega_1=\frac{\phi}{L}$,
$\omega_2=\frac{2\pi l+\phi}{L}$ and
$\omega_3=\frac{2\pi l-\phi}{L}$,
and $l_m$, $l_n$ are the largest integer satisfying
$\omega_{2,3}<\mu$, respectively and if $\omega_{2,3}>\mu$,
then $\Omega_{\beta}=0$.

It is interesting to study the case that
$\mu$ is less than, or at most nearly equals to
$2\pi/L$. We only, therefore, consider $0<\mu L<\pi$ case.

In $M_3\times S^1$ space-time,
one can express the analytical form:
\begin{equation}
 \Omega_{\beta}
 =
 -\frac{N_f}{3}\frac{1}{\pi L^4}
 (\phi-\mu L)^2
 (2\phi+\mu L),
 \label{omegabeta2}
\end{equation}
for $0\leq\phi\leq\mu L$,
\begin{equation}
 \Omega_{\beta}=0,
\end{equation}
for $\mu L<\phi\leq 2\pi-\mu L$,
\begin{equation}
 \Omega_{\beta}
 =-\frac{N_f}{3}\frac{1}{\pi L^4}
 \left(
 2\pi-\phi-\mu L
 \right)^2
 (4\pi-2\phi+\mu L),
\end{equation}
for $2\pi-\mu L<\phi<2\pi$.

Further, in $M_4\times S^1$ space-time, one can also represent
the analytical form:
\begin{equation}
 \Omega_{\beta}
 =-
 \frac{N_f}{12\pi^2L^5}
 \left[
 \mu L(2\mu^2L^2-5\phi^2)
 \sqrt{\mu^2L^2-\phi^2}
 +3\phi^4
 \ln\left(
 \frac{\mu L}{\phi}
 +\sqrt{\frac{\mu^2L^2}{\phi^2}-1}
 \right)
 \right],
 \label{omegabeta3}
\end{equation}
for $0\leq\phi\leq\mu L$,
\begin{equation}
 \Omega_{\beta}=0,
\end{equation}
for $\mu L<\phi\leq 2\pi-\mu L$,
\begin{eqnarray}
 \Omega_{\beta}
 =
 &-&
 \frac{N_f}{12\pi^2L^5}
 \left[
 \mu L
 \{
 2\mu^2L^2-5(2\pi-\phi)^2
 \}
 \sqrt{\mu^2L^2-(2\pi-\phi)^2}
 \right.\nonumber \\
 & &
 \left.
 +3\phi^4
 \ln\left(
 \frac{\mu L}{2\pi-\phi}
 +\sqrt{\frac{\mu^2L^2}{(2\pi-\phi)^2}-1}
 \right)
 \right],
\end{eqnarray}
for $2\pi-\mu L<\phi<2\pi$.
\section{construction of a vacuum gauge ball}
With the expressions of the thermodynamic potential
in the presence of a strongly degenerate fermion gas
in $M_3\times S^1$ and $M_4\times S^1$ space-times
in the previous section, we will show that
if the configuration of $\phi$ is a domain wall
configuration, one can construct a vacuum gauge ball,
observing that the thermodynamic potential
has a barrier between their vacua, 
and the two vacua are almost degenerate with the energy
density difference, which is written as a function of $\mu L$,
in the case of $0\leq\phi\leq\mu L$.
\subsection{$M_3\times S^1$ space-time}
We rewrite the thermodynamic potential,
from (\ref{omegazero2}) and (\ref{omegabeta2}),
in the case of $0\leq\phi\leq\pi$,
\begin{equation}
 \Omega
 =
 \frac{N_f}{6\pi^2L^4}
 \left[
 2\pi^2(\phi-\pi)^2-(\phi-\pi)^4
 \right]
 -\frac{N_f}{3\pi L^4}
 (\phi-\mu L)^2(2\phi+\mu L)
 \theta(\mu L-\phi).
 \label{2d}
\end{equation}

In Fig. 1, we show the thermodynamic potential for various
values of $\mu L$.
At $\phi=0$, $\pi$, the values of the thermodynamic potential
are given by
\begin{eqnarray}
 \Omega&=&
 \frac{N_f\pi^2}{6L^4}-\frac{N_f(\mu L)^3}{3\pi L^4},
 \qquad
 \mbox{at} \quad \phi=0,
 \\
 \Omega&=&
 0,
 \qquad
 \mbox{at} \quad \phi=\pi,
\end{eqnarray}
respectively.
There are three cases, which lead to the distinct
physical vacua for values of $\mu L$.
For $0\leq\mu L<2^{-3/2}\pi$ case,
the thermodynamic potential has a minimum at $\phi=\pi$ only.
For $\mu L=2^{-3/2}\pi$ case,
the values of the thermodynamic potential
at $\phi=0$ and $\phi=\pi$ are degenerate.
For $\mu L>2^{-3/2}\pi$ case,
the thermodynamic potential has a minimum at $\phi=0$.
If one considers the case of $\mu L\gtrsim 2^{-3/2}\pi$,
the two vacua (at $\phi=0$ and $\phi=\pi$) become almost degenerate.
Within the viewpoint of symmetry breaking \cite{ref},
the fermions, which are massless at the classical level,
acquire dynamical masses through quantum corrections
for $0\leq\mu L<2^{-3/2}\pi$ case,
because the physical vacuum has a minimum at 
the non-trivial configuration $\phi=\pi$.
Moreover, for $\mu L>2^{-3/2}\pi$ case,
the dynamical masses of the fermions disappear
because the physical vacuum is at the trivial configuration
of $\phi=0$.
In general, for non-abelian gauge theory, the different vacua
correspond to breaking and restoration of gauge symmetry \cite{ref3}.
Thus, for degenerate fermions, the finite density effects
are crucial to investigate whether symmetry is breaking or not.

Now, in order to construct a vacuum gauge ball,
we restrict our attention to the case of the almost degenerate
vacua ($\mu L\gtrsim 2^{-3/2}\pi$).
We should assume that the space-time has a spherical symmetry
and further the vacuum gauge ball does not have the $U(1)$ charge,
otherwise, the Coulomb force between
fermions destroys the strong degenerate state.
The situation we assume is whether the gauge coupling constant is
very small, or the charge of each fermion, which composes
the vacuum gauge ball, compensates with each other.
Thus, $\phi$ is a function of the radial coordinate $r$ only
and the Maxwell term in the Lagrangian (\ref{lag})
will be reduced to
\begin{equation}
 -\frac{1}{4}F^{\mu\nu}F_{\mu\nu}=-\frac{1}{2}
 \frac{1}{g^2L^2}(\phi'(r))^2,
\end{equation}
where $'$ denotes the derivative with respect to $r$. 

For the gauge field, the equation of motion takes the form
\begin{equation}
 \frac{1}{r}(r\phi')'-
 g^2L^2\frac{\partial\Omega}{\partial\phi}
 =0,
 \label{vgb}
\end{equation}
where we have assumed that $L$ is fixed to a constant value.

From (\ref{2d}), thermodynamical quantities
at $\beta\to\infty$ are as follows
\begin{eqnarray}
 n(r)
 &=&
 -\frac{\partial\Omega}{\partial\mu}
 \Bigg|_{V_2,L}
 =\frac{N_f}{\pi L^3}
 (\mu^2L^2-\phi^2)\theta(\mu L-\phi),
 \\
 \rho(r)
 &=&
 \Omega+\mu n(r)+\frac{1}{2}\frac{1}{g^2L^2}(\phi')^2,
 \label{energydensity}
 \\
 &=&
 \frac{N_f}{3\pi L^4}
 \left[
 2(\mu^3L^3-\phi^3)\theta(\mu L-\phi)
 +
 \pi(\phi-\pi)^2-\frac{1}{2\pi}(\phi-\pi)^4
 \right]
 +\frac{1}{2}\frac{1}{g^2L^2}
 (\phi')^2,
\end{eqnarray}
where $\rho(r)$ and $n(r)$ are the energy density and
the particle number density, respectively.
In (\ref{energydensity}), the first and second terms in the right
hand side are the contributions from the fermions (coupled with the
vacuum gauge field),
and the third term is that from the vacuum gauge field only
to the energy density.

By analyzing the field equation (\ref{vgb}) numerically,
one can search for the configuration of $\phi(r)$, which is
the case of a domain wall configuration, as shown in Fig. 2,
for a positive value of $\mu L$ ($\gtrsim 2^{-3/2}\pi$).
In this configuration, we show the profiles of the energy density
$\rho(r)$ and the particle number density $n(r)$ of
the vacuum gauge ball, in Fig. 3.
Then, one can obtain the physical values
(the mass $M$ and the particle number $N$)
of the vacuum gauge ball as follows
\begin{eqnarray}
 M&=&
 2\pi L
 \int^{\infty}_0
 \rho(r) r dr, \\
 N&=&
 2\pi L
 \int^R_0
 n(r) r dr,
\end{eqnarray}
where $R$ is a radius of the vacuum gauge ball,
and $n(r)=0$ and $\rho(r)=0$ for $r>R$.

For the various positive values of $\mu L$ ($\gtrsim 2^{-3/2}\pi$),
we construct the vacuum gauge balls and
plot the relationship between the mass $\tilde{M}$
and the particle number $\tilde{N}$
($\tilde{M}=g^2LM$ and $\tilde{N}=g^2N$) in Fig. 4.
As far as we can form them numerically,
one sees that they are stable energetically
because of satisfying the condition of
(mass $\times$ particle number) $>$ (energy),
where the mass (of the fermion in the vacuum ) is $\pi/L$,
and the dots representing the ball should be connected
to the straight line $\tilde{M}=\pi\tilde{N}$,
which is similar to the NTS case \cite{NTS}.

In Fig. 5, we show the relation between the particle number
$\tilde{N}$ and the radius $R$.
One sees that the larger vacuum gauge balls are formed
as the particle numbers increase.
\subsection{$M_4\times S^1$ space-time}
One can perform the construction of the vacuum gauge ball,
using the manners in $M_3\times S^1$ space-time.
We rewrite the thermodynamic potential,
from (\ref{omegazero}) and (\ref{omegabeta3}),
in the case of $0\leq\phi\leq\pi$,
\begin{eqnarray}
 \Omega
 &=&
 \frac{6N_f}{\pi^2L^5}
 \sum^{\infty}_{l=1}\frac{\cos(l\phi)-(-1)^l}{l^5}
 -
 \frac{N_f}{12\pi^2L^5}
 \left[
 \mu L(2\mu^2L^2-5\phi^2)
 \sqrt{\mu^2L^2-\phi^2}
 \right. \nonumber \\
 & &
 \left.
 +3\phi^4
 \ln\left(
 \frac{\mu L}{\phi}
 +\sqrt{\frac{\mu^2L^2}{\phi^2}-1}
 \right)
 \right]
 \theta(\mu L-\phi).
 \label{omegabeta}
\end{eqnarray}
At $\phi=0$, $\pi$, the values of the thermodynamic
potential are obtained by
\begin{eqnarray}
 \Omega&=&
 \frac{6N_f}{\pi^2L^5}
 \sum^{\infty}_{l=1}\frac{1-(-1)^l}{l^5}
 -\frac{N_f(\mu L)^4}{6\pi^2 L^5},
 \qquad
 \mbox{at} \quad \phi=0,
 \\
 \Omega&=&
 0,
 \qquad
 \mbox{at} \quad \phi=\pi,
\end{eqnarray}
respectively. There are three cases,
which lead to the different physical vacua for values of $\mu L$
in such case as $M_3\times S^1$ space-time.
At $\mu L=(\frac{279}{4}\zeta(5))^{1/4}\simeq 2.91624$,
the two vacua are degenerate, where we have used the facts of
$\sum^{\infty}_{l=1}\frac{1}{l^5}=\zeta(5)$ and
$\sum^{\infty}_{l=1}\frac{(-1)^l}{l^5}=-\frac{15}{16}\zeta(5)$.

For the vacuum gauge ball, we should take the same assumptions
for the symmetry of the space-time and the global charge
of the ball.
Then, the equation of motion for
the vacuum gauge field with the fixed value of $L$ is given by
\begin{equation}
 \frac{1}{r^2}(r^2\phi')'-
 g^2L^2\frac{\partial\Omega}{\partial\phi}=0.
 \label{41}
\end{equation}
From (\ref{omegabeta}), thermodynamic quantities are
as follows
\begin{eqnarray}
 n(r)
 &=&
 \frac{2N_f}{3\pi^2 L^4}
 (\mu^2L^2-\phi^2)^{3/2}
 \theta(\mu L-\phi),
 \\
 \rho(r)
 &=&
 \frac{6N_f}{\pi^2L^5}\sum^{\infty}_{l=1}
 \frac{\cos(l\phi)-(-1)^l}{l^5}
 +\frac{N_f}{4\pi^2L^5}
 \left[
 \mu L(2\mu^2L^2-\phi^2)
 \sqrt{\mu^2L^2-\phi^2}
 \right. \nonumber \\
 & & \left.
 -\phi^4
 \ln
 \left(
 \frac{\mu L}{\phi}
 +
 \sqrt{\frac{\mu^2L^2}{\phi^2}-1}
 \right)
 \right]
 \theta(\mu L-\phi)
 +\frac{1}{2}\frac{1}{g^2L^2}(\phi')^2.
 \label{48}
\end{eqnarray}

By solving the field equation (\ref{41}) numerically,
one can search for a domain wall configuration for $\phi$,
for a positive value of $\mu L$
$(\gtrsim(\frac{279}{4}\zeta(5))^{1/4})$,
appeared in like Fig. 2.
In this configuration, one can obtain the physical values
of the vacuum gauge ball as follows
\begin{eqnarray}
 M&=&
 4\pi L\int^{\infty}_0
 \rho(r) r^2 dr,
 \\
 N&=&
 4\pi L
 \int^{R}_0
 n(r) r^2 dr,
\end{eqnarray}
where $R$ is a radius of the vacuum gauge ball,
and $n(r)=0$ and $\rho(r)=0$ for $r>R$.

For the various positive values of $\mu L$
$(\gtrsim(\frac{279}{4}\zeta(5))^{1/4})$,
we also construct the vacuum gauge ball and
display the relationship between the mass $\tilde{M}$
and the particle number $\tilde{N}$ in Fig. 6,
where 
\begin{equation}
 \tilde{M}=\sqrt{\frac{N_f}{6\pi^2}}
 \left(\frac{g}{L^{\frac{1}{2}}}\right)
 g^2M,
 \qquad
 \tilde{N}=\sqrt{\frac{N_f}{6\pi^2}}
 \left(\frac{g}{L^{\frac{1}{2}}}\right)^3
 N.
\end{equation}
$\frac{g}{L^{\frac{1}{2}}}$ is reduced to
the effective four-dimensional gauge coupling.
Unlike the case of $M_3\times S^1$ space-time, we see that
the dots showing the ball solutions should be connected
to the straight line $\tilde{M}=\pi\tilde{N}$,
but for some small masses of the balls, the dots cross the line.
This situation is also similar to the NTS case.
For some small particle number $\tilde{N}$,
the balls seem to be energetically unstable,
but the balls will be energetically stable
since we assume that $N_f$ and $N$ are macroscopic quantity
so that $\tilde{N}$ are also macroscopic quantity.
And then the values of $\tilde{M}$ are large for macroscopic
quantity of $\tilde{N}$ so that the mass of the balls will
be heavy.

Further, we plot the relation between the particle number $\tilde{N}$
and the radius $R$ in Fig. 7.
One sees that the radius has a minimum value for the non-zero
particle number. The reason we think is that if the particle number
is small, the contribution from the surrounding wall is larger
than that from the fermions to the total energy of the
vacuum gauge ball.
The actual radius of the ball is given by
\begin{equation}
 \sqrt{\frac{6\pi^2}{N_f}}
 \left(\frac{L^{\frac{1}{2}}}{g}\right)
 L R.
\end{equation}
Even if the large values of $R$ are obtained for macroscopic
quantity of $\tilde{N}$, form Fig. 7,
the circumference of $S^1$ is set to
inverse of some energy scale, for example,
$L\sim {\rm (TeV)^{-1}}$
so that the actual size of the balls will be small.
Thus, this small but heavy ball, which does not have the $U(1)$
charge, will be a candidate for unknown matter
such as Q-ball, NTS and F-ball.
\section{discussion}
In this paper, we have considered the abelian gauge theory
coupled with the Dirac fermion
in $M_3\times S^1$ and $M_4\times S^1$ space-times.
We have calculated the thermodynamic potential for the vacuum
gauge field in the presence of a strongly degenerate fermion gas,
which is parametrized as $\mu L$.

We have constructed the vacuum gauge ball,
observing the three conditions:
the configuration of $\phi$ is a domain wall configuration,
the thermodynamic potential has a barrier between their vacua
and the two vacua are almost degenerate for some values of $\mu L$.

Performing the numerical analysis of the field equation
for the vacuum gauge field,
we have found the relationship between the mass $\tilde{M}$ and
the particle number $\tilde{N}$ for various vacuum gauge balls
in Fig. 4 and Fig. 6, and between the radius $R$ and
the particle number $\tilde{N}$ in Fig. 5 and Fig. 7.
For the behavior between $\tilde{M}$ and $\tilde{N}$,
the situation we have obtained is similar to the NTS case.
But for the behavior between $R$ and $\tilde{N}$
in $M_4\times S^1$ space-time,
the situation is different from that in the NTS case.
The reason we think is that when the particle number is small,
the contribution from the surrounding wall becomes larger than
that from the fermions to the total energy of the vacuum gauge ball.
It is still unclear that the behavior between $\tilde{N}$ and $R$
for some small values of $\tilde{N}$ differs
in $M_3\times S^1$ and $M_4\times S^1$ space-times.

Although we have not investigated the cases of
non-abelian gauge theory and other topology of extra dimensions,
the vacuum gauge balls we have constructed may be sufficient to
know the fundamental properties of the ball,
which is constructed from more complicated situations.

So far, our analysis is confined to the flat background space-time.
It would also be interesting to consider the vacuum gauge ball in
curved space-time, such as Q-star \cite{Qstar} and
Boson star (for a review, see Ref. \cite{Bosonstar}),
and to construct the ball in the little Higgs model
(or a model with dimensional deconstruction \cite{DD}).
Further, in order to investigate the stability of the vacuum
gauge balls, the calculation of quantum correction to
the their energy is very important.
We must continue to make every effort to study these
situations.
\begin{figure}[h]
\epsfxsize=7cm
\mbox{\epsfbox{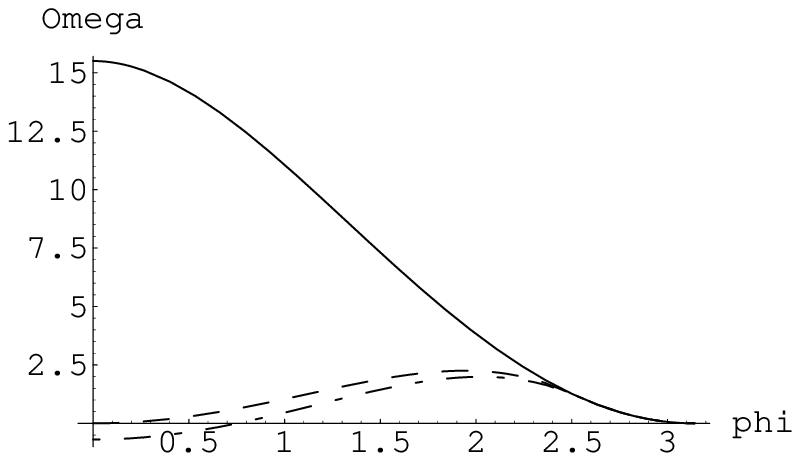}}\\
\bigskip
\caption{The thermodynamic potential $\Omega$ of
(\ref{2d}) is displayed against $\phi$ for various values of
$\mu L$ in $M_3\times S^1$ space-time.
The solid line corresponds to $\mu L=0$, the dashed line
corresponds to $\mu L=2^{-3/2}\pi$ (two degenerate vacuum case),
and the dot-dashed line corresponds to
$\mu L=2.5$ $(\gtrsim2^{-3/2}\pi)$
(two almost degenerate vacuum case).}
\label{fig1}
\epsfxsize=7cm
\mbox{\epsfbox{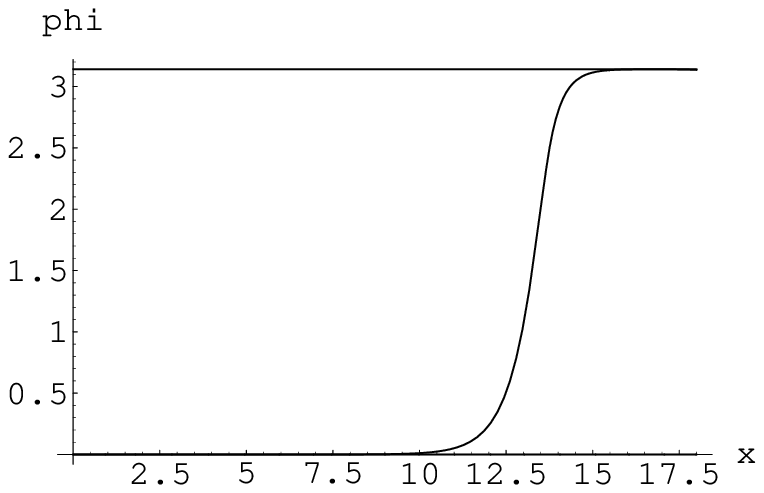}}\\
\bigskip
\caption{The configuration of
$\phi(x)$, which is a domain wall configuration,
is shown against $x$ ($=\sqrt{\frac{N_f}{3\pi}}\frac{gr}{L}$).}
\epsfxsize=7cm
\mbox{\epsfbox{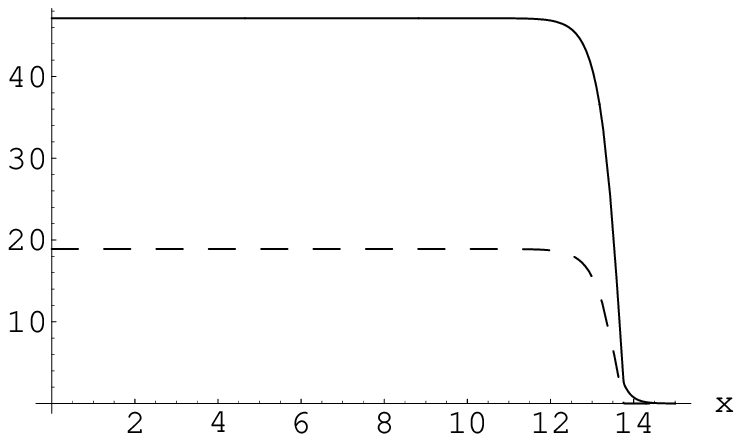}}\\
\bigskip
\caption{
The profiles of the energy density $\rho(x)$ and
the particle number density $n(x)$ are shown against $x$
($=\sqrt{\frac{N_f}{3\pi}}\frac{gr}{L}$)
in the case that the configuration of $\phi$
is a domain wall configuration.
The solid line corresponds to $\rho(x)$ and the dashed line
corresponds to $n(x)$.
}
\end{figure}
\begin{figure}[h]
\epsfxsize=8cm
\mbox{\epsfbox{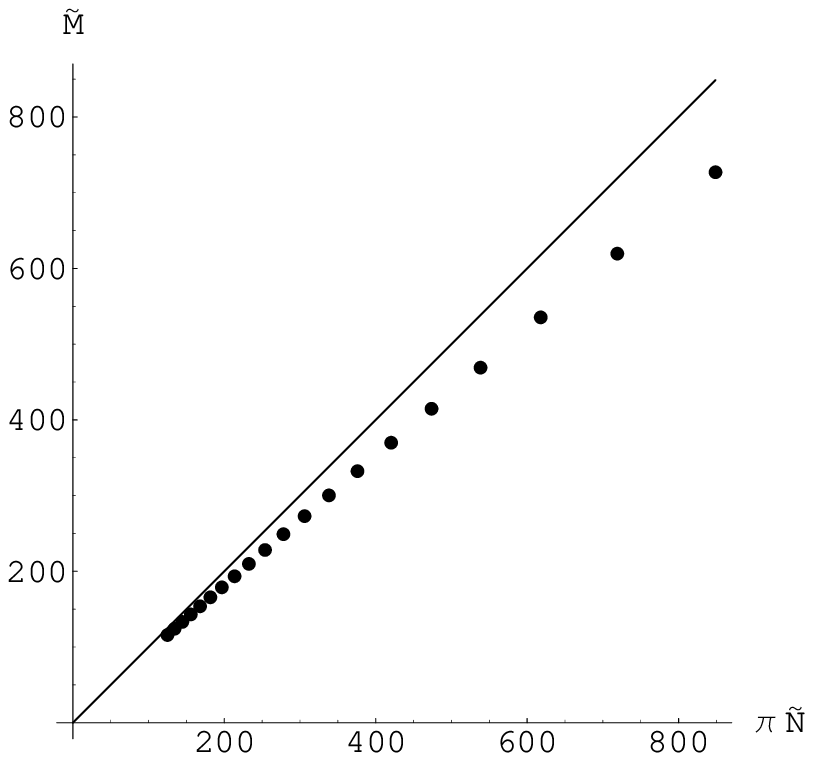}}\\
\bigskip
\caption{
For the various positive values of $\mu L$ ($\gtrsim 2^{-3/2}\pi$),
the relationship between the mass $\tilde{M}$
and the particle number $\tilde{N}$ of the vacuum gauge ball
is shown, where $\tilde{M}$
and $\tilde{N}$ are defined as $\tilde{M}\equiv g^2 L M$
and $\tilde{N}\equiv g^2 N$, respectively
in $M_3\times S^1$ space-time.
The straight line corresponds to $\tilde{M}=\pi\tilde{N}$.
}
\epsfxsize=10cm
\mbox{\epsfbox{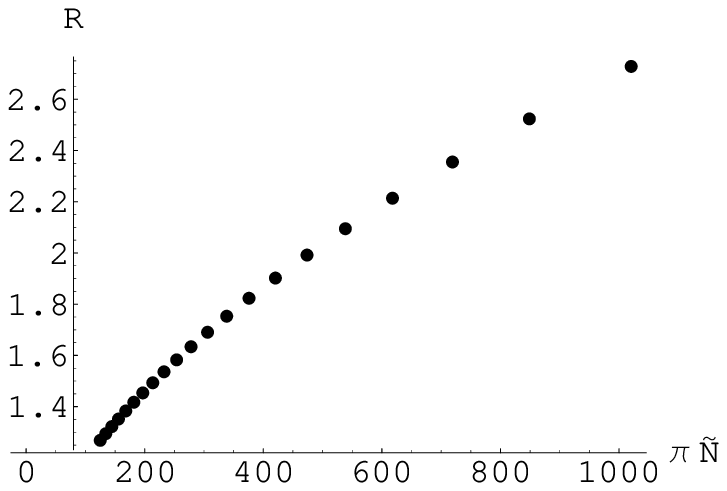}}\\
\bigskip
\caption{
For the various positive values of $\mu L$ ($\gtrsim 2^{-3/2}\pi$),
the relationship between the radius $R$ and the particle number
$\tilde{N}$ for the vacuum gauge ball is shown
in $M_3\times S^1$ space-time.
}
\end{figure}
\pagebreak
\begin{figure}[h]
\epsfxsize=8cm
\mbox{\epsfbox{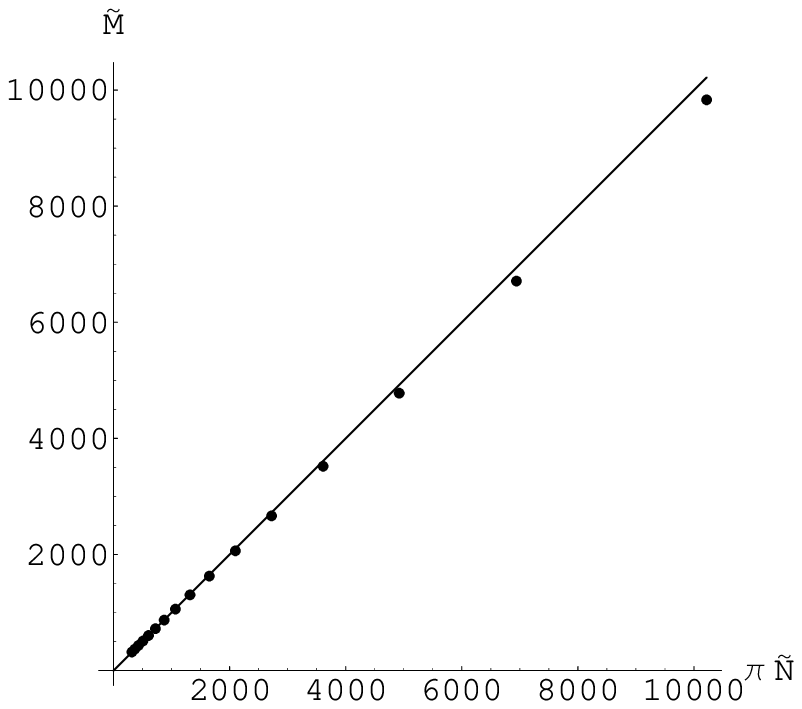}}\\
\bigskip
\caption{
For the various positive values of $\mu L$
$(\gtrsim (\frac{279}{4}\zeta(5))^{1/4})$,
the relationship between the mass $\tilde{M}$
and the particle number $\tilde{N}$ for the vacuum gauge ball
is shown, where $\tilde{M}$ and $\tilde{N}$ are defined as
$\tilde{M}\equiv\frac{g^3}{\pi}\sqrt{\frac{N_f}{6L}}M$
and $\tilde{N}\equiv\frac{g^3}{\pi L^{3/2}}\sqrt{\frac{N_f}{6}}N$,
respectively in $M_4\times S^1$ space-time.
The straight line corresponds to $\tilde{M}=\pi\tilde{N}$.
}
\epsfxsize=8cm
\mbox{\epsfbox{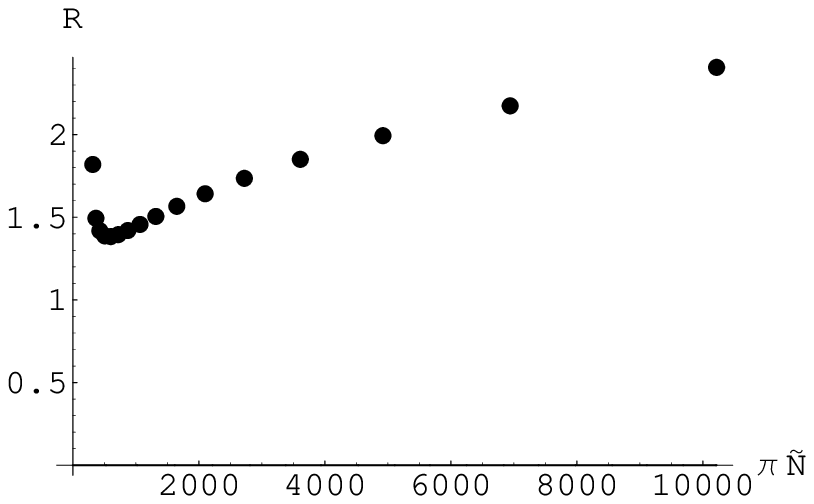}}\\
\bigskip
\caption{
For the various positive values of $\mu L$
$(\gtrsim (\frac{279}{4}\zeta(5))^{1/4})$,
the relationship between the radius $R$ and the particle number
$\tilde{N}$ for the vacuum gauge ball is shown
in $M_4\times S^1$ space-time.
}
\end{figure}
\begin{acknowledgments}
We would like to thank N. Kan and R. Takakura
for their valuable comments and for reading the manuscript.
\end{acknowledgments}

\end{document}